\documentclass[affil-it, twocolumn]{article}
\usepackage{authblk}
\usepackage{amsmath}
\usepackage{subfigure}
\usepackage{graphicx}			
\usepackage{epsfig} 			
\usepackage{mathptmx} 			
\usepackage{amsfonts}
\usepackage{amssymb}
\usepackage{epstopdf}
\usepackage[english]{babel}
\usepackage{pgfplots}
\usepackage{url}
\usepackage[export]{adjustbox}
\usepackage[labelfont=bf]{caption}
\usepackage{titlesec}
\usepackage{multirow}
\usepackage{tabularx}
\usepackage{array}

\titleformat{\subsection}[runin]
{\normalfont\normalsize\bfseries}{\thesubsection}{1em}{}
 \def\ELEGANT { \hskip -4pt {\scshape elegant} }

\begin{document}
\title{Generation and Measurement of Sub-Micrometer Relativistic Electron Beams}
\author[1]{Simona Borrelli \thanks{simona.borrelli@psi.ch}}
\author[1]{Gian Luca Orlandi}
\author[1]{Martin Bednarzik}
\author[1]{Christian David}
\author[2]{Eugenio Ferrari}
\author[1]{Vitaliy A. Guzenko}
\author[1]{Cigdem Ozkan-Loch}
\author[1]{Eduard Prat}
\author[1]{Rasmus Ischebeck}
\affil[1]{Paul Scherrer Institut, 5232 Villigen PSI, Switzerland}
\affil[2]{\'Ecole Polytechnique F\'ed\'erale de Lausanne EPFL, Lausanne, Switzerland}

\date{\today}

\twocolumn[
	\begin{@twocolumnfalse}
		\maketitle
		\setcounter{secnumdepth}{0}
		\renewcommand{\abstractname}{}
		\begin{abstract}
		The generation of low-emittance electron beams has received significant interest in recent years. Driven by the requirements of X-ray free electron lasers, the emittance of photocathode injectors has been reduced significantly, with a corresponding increase in beam brightness. At the same time, this has put increasingly stringent requirements on the instrumentation to measure the beam size.  These requirements are even more stringent for novel accelerator developments, such as laser-driven accelerators based on dielectric structures or on a plasma, or for linear colliders at the energy frontier.  We present here the generation and measurement of a sub-micrometer electron beam, at a particle energy of 330 MeV, and a bunch charge below 1 pC.  An electron beam optics with a $\beta$-function of a few millimeters in the vertical plane had been set up.  The beam was characterized through a wire scanner that employs a 1 $\mathrm{\mu m}$ wide metallic structure fabricated using the electron beam lithography on a silicon nitride membrane. The smallest (rms) beam size presented here is less than 500 nm.
		\end{abstract}
	\end{@twocolumnfalse}
]
\clearpage
Nowadays, scientists are striving for electron beams with extremely small transverse sizes and emittances, driven by the high demand of X-ray free-electron-lasers (XFELs) and novel accelerators development. 
In linear colliders one of the main design goals is to achieve a small beam size at the interaction point to maximize the luminosity. 
The design of the International Linear Collider, for example, uses a nanometer-sized beam at the interaction point \cite{behnke2013international, balakin1995focusing}.
Similar requirements have to be met for advanced accelerator concepts, such as dielectric laser accelerators \cite{england2014dielectric, peralta2013demonstration} and plasma wake-field accelerators \cite{blumenfeld2007energy,sears2010emittance, weingartner2012ultralow}. These schemes utilize micron scale transverse accelerating structures, defined by either the plasma or laser wavelength, and have been shown to achieve high accelerating gradients. 
In a dielectric laser accelerator, the electron beam is accelerated through a dielectric microstructure that requires beam transverse dimensions in the sub-$\mathrm{\mu}$m scale \cite{wootton2016dielectric}. 
In the field of plasma based acceleration, reducing the transverse emittance down to the nanometer scale would open a path towards many different applications of this technology, such as colliders and light sources \cite{xu2014low}.
Concerning XFELs the reduced beam emittance would allow for a compact design with lower beam energy and shorter undulator length \cite{rosenzweig2016next, milne2017swissfel, prat2014emittance}.

The challenge of designing ultra-low emittance accelerators brings the need for transverse beam size diagnostics with sub-micrometer resolution.
However, the resolution limit of commonly used beam profile monitors hinders the possibility to measure sub-micrometer spot sizes \cite{tenenbaum1999measurement,ischebeck2015transverse,orlandi2016design} (details are provided in 'Discussion').

\begin{figure}[h]
	\centering
	\includegraphics[width = 0.5\textwidth]{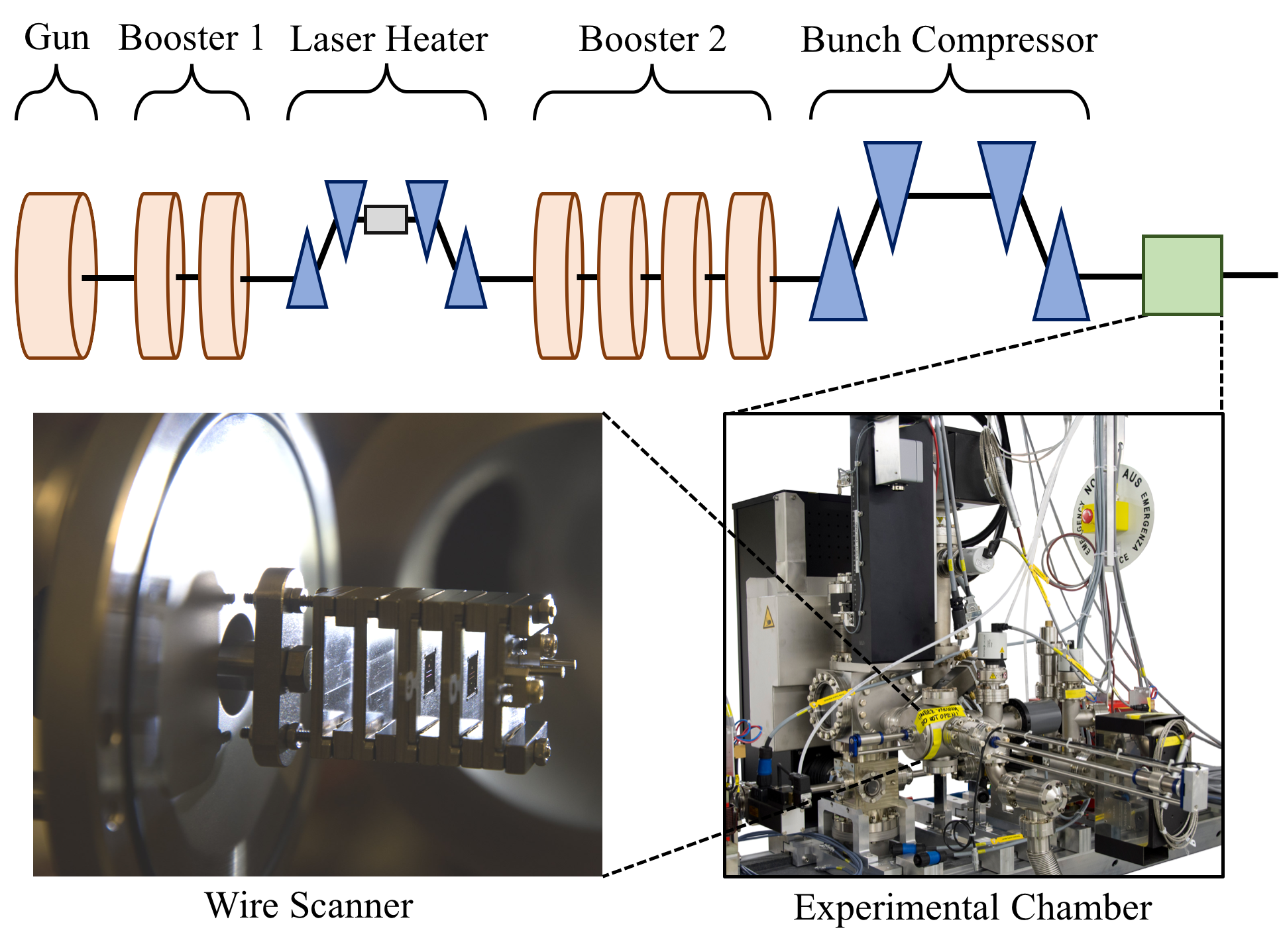}
	\caption{\textbf{Layout of the SwissFEL injector.} The electron beam is emitted in an RF photoinjector and then accelerated up to an energy of 330 MeV by an S-band RF booster linac (Booster 1 and 2). An experimental chamber is located downstream of a bunch compressor and offers the possibility to insert into the beamline the designed wire scanner on-a-chip. We show a picture of the experimental chamber (bottom right) and the wire scanners installed on sample holders in the load-lock pre-chamber (bottom left). \label{fig:layout_swissfel}}
\end{figure}

We designed a novel wire scanner (WSC hereinafter) with resolution in the sub-micrometer range. 
In a WSC measurement a thin metallic wire scans the beam transversally. The generated particle shower (loss signal) is detected downstream, enabling the reconstruction of the beam transverse profile (see 'Methods'). The design of a WSC with improved resolution requires a reduction of the wire width.
We developed a high-resolution profile monitor exploiting nanofabrication techniques to produce a 1 $\mathrm{\mu}$m wide metallic stripe on a membrane by electron beam lithography and electroplating.
The proposed design is a break-through in the state-of-the-art of transverse profile monitors ensuring sub-micrometer resolution as well as enabling the integration of a wire scanner on-a-chip.
We present in this paper the generation and characterization of an electron beam with a 53 nm normalized emittance in the vertical plane, for a particle energy of 330 MeV and a bunch charge below 1 pC.
This nanometer emittance has been attained at SwissFEL \cite{milne2017swissfel, ganter2010swissfel}, the X-ray free electron laser at the Paul Scherrer Institute. 
We achieved sub-micrometer spot sizes setting up an electron beam optics to meet a 2.61 mm $\beta$-function in the vertical plane. 
The presented acceleration setup, beam optics and wire scanner on-a-chip, have enabled the generation and measurement of beams with rms sizes less than 500 nm.
To the best of the author's knowledge, this is the smallest relativistic electron beam measured at an FEL accelerator.

\section{Results}\label{sec:Results}
\subsection{Generation of sub-micrometer relativistic electron beams. }
Figure \ref{fig:layout_swissfel} shows the layout of the SwissFEL injector, where we generated and characterized the relativistic electron beams with rms transverse sizes below 500 nm. 

In the SwissFEL injector, electron beams are emitted in an RF photoinjector gun \cite{Raguin:2013txa} and initially accelerated in an S-band RF booster linac (consisting of two sections: Booster 1 and 2) up to a maximum energy of 330 MeV \cite{raguin2012swiss}. 
In the last injector section, 
an experimental chamber \cite{ferrari2017achip} located downstream of a bunch compressor, offers the possibility of scanning the novel wire scanner on-a-chip vertically through the beam to measure the vertical beam profile.

The beam size was brought down to sub-micrometer scale by minimizing the beam emittance at the photoinjector and the vertical $\beta$-function at the interaction point with the wire scanner (see section 'Methods').
The emittance was minimized by suitably reducing the laser spot size on the cathode and by operating the gun at a very low charge (below 1 pC) where space-charge effects are negligible. The measured vertical normalized emittance was 53 nm, with estimated errors of $\sim 10 \%$.

Once the gun is optimized for the smallest emittance, the achievement of a sub-micrometer size hinges on the minimization of the vertical $\beta$-function.
For this purpose, we calculated and implemented in the accelerator a specifically suited optics. 
The computed vertical $\beta$-function at the wire scanner location was 2.61 mm.

Using an electron energy of 330 MeV and the presented vertical normalized emittance and $\beta$-function, we were able to generate a beam with an expected vertical size of $460 \pm 20$ nm at the wire-scanner location. 

\subsection{Beam profile measurements with sub-micrometer resolution.}
The goal of measuring sub-micrometer beam sizes can not be achieved using conventional profile monitors such as screens and wire scanners, since their resolution is typically limited to the micrometer scale (see section 'Discussion'). Among them wire scanners are highly promising devices in terms of spatial resolution, which mainly depends on the wire width. 
In the conventional design this dimension is constrained to a few micrometers \cite{orlandi2017commissioning} (see section 'Methods').
To overcome this limitation, we exploited the possibilities offered by nanofabrication techniques to develop a novel WSC on-a-chip with improved sub-micrometer resolution. 
The device we present here consists of a silicon chip with a silicon nitride membrane at the center (cf. Fig. \ref{fig:wsc_on_a_chip}). On the membrane two gold stripes are produced by electron-beam lithography and electroplating. They are characterized by different widths, namely $w_{1} = 1 \; \mathrm{\mu m}$ and $w_{2} = 2 \; \mathrm{\mu m}$. Consequently, their rms geometrical resolutions are 0.3 and 0.6 $\mathrm{\mu}$m respectively (see 'Methods'). 
A vertical scan through the beam is performed for each stripe
independently to measure the beam profile. 
For comparison, we also scan through the beam a conventional cylindrical 5 $\mathrm{\mu}$m tungsten wire (cf. Fig. \ref{fig:all_prototypes}) whose rms geometrical resolution is 1.25 $\mathrm{\mu}$m \cite{orlandi2016design}. 

\begin{figure}[]
	\centering
	\includegraphics[width = 1\columnwidth]{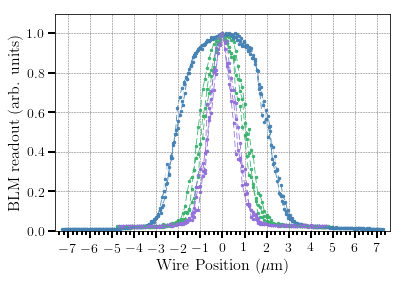}
	\caption{\textbf{Wire scanner beam profile measurements.} Beam vertical profile measurements performed scanning through the beam a 5 $\mathrm{\mu m}$ tungsten wire (in blue), a 2 $\mathrm{\mu m}$ (in green) and a 1 $\mathrm{\mu m}$ (in purple) gold stripe. For each wire, we show three sequential and suitably normalized beam profile measurements.\label{fig:Global_plot}} 
\end{figure}

In Fig. \ref{fig:Global_plot} we present measurements of the beam vertical profile obtained scanning each of the three wires through the beam and acquiring the corresponding loss signal by a beam loss monitor (BLM). 
For each wire, we show three sequential beam profile acquisitions. 
We evaluated the rms size of each of the three profiles. In Table \ref{table:0} we report their mean values $\sigma_{\mathrm{rms}}$, with their associated statistical errors. 
Only the $\sigma_{\mathrm{rms}}$ from the 1 $\mathrm{\mu m}$ gold stripe measurements is comparable with the expected beam dimension of $460\pm 20$ nm. This behavior can be attributed to the resolution limits of the 5 $\mathrm{\mu m}$ tungsten wire and the 2 $\mathrm{\mu m}$ gold stripe.
The measured traces are indeed the convolution of the beam profile with the wire cross section. The cylindrical shape of the 5 $\mathrm{\mu m}$ wire is clearly visible in Fig. \ref{fig:Global_plot}.

\begin{figure}[]
	\centering
	
	\subfigure[]{
		\includegraphics[ trim={0 8 0 5},clip,height = 0.29\textwidth]{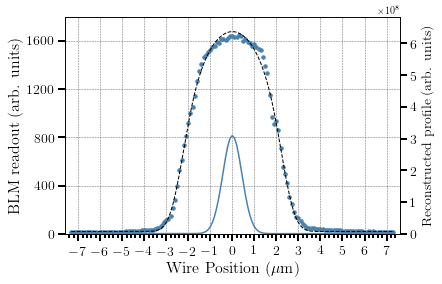}\label{fig:Conv_profile_center_5um_1}
	} 
	\subfigure[]{
		\includegraphics[ trim={0 8 0 5},clip,height = 0.29\textwidth]{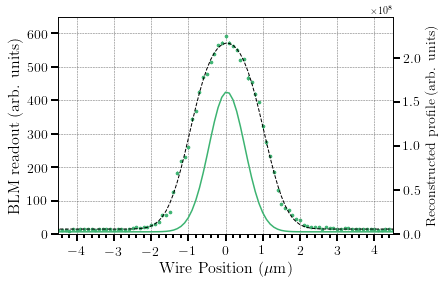}\label{fig:Conv_profile_center_2um_5}
	} 
	\subfigure[]{
		\includegraphics[ trim={0 8 0 5},clip,height = 0.29\textwidth]{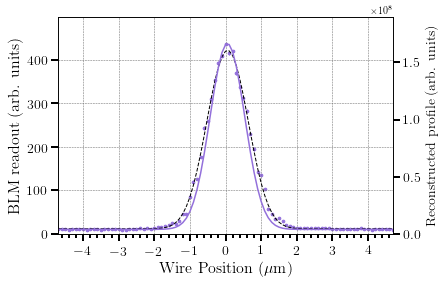}\label{fig:Conv_profile_center_1um_5} 
	}\vspace{-0.25cm}
	\caption{\textbf{Beam vertical size reconstructions from wire scanner measurements.} One of the three profiles measured scanning through the beam (a) the conventional 5 $\mathrm{\mu}$m W wire and the novel (b) 2 $\mathrm{\mu}$m and (c) 1 $\mathrm{\mu}$m Au stripes. We show in dotted black line the fit curve and in solid colored line the reconstructed Gaussian beam profile.  
	Refer to Table \ref{table:0} for the corresponding beam sizes.  \label{fig:BeamProfiles}}
\end{figure}

Nevertheless, the beam size can be extracted from the measured profiles fitting the data with the convolution of a function describing the wire shape and one describing the beam, which we assume to be Gaussian (details are provided in section 'Methods').
Figure \ref{fig:BeamProfiles} shows one of the three beam profiles measured by means of the conventional 5 $\mathrm{\mu}$m W wire, the novel 2 and 1 $\mathrm{\mu}$m gold stripes, and the corresponding fit curves. It also shows the reconstructed Gaussian profiles obtained by the described deconvolution.
The profile measured by the 1 $\mathrm{\mu}$m gold stripe and the corresponding reconstructed Gaussian profile are comparable. This confirms the goodness of our initial assumption of a Gaussian beam shape.

The beam vertical sizes $\sigma_\mathrm{y}$ derived from the fit procedure are summarized in Table \ref{table:0}. 
Each value is the average of the fit sigma over the three sequential beam profile measurements, with the corresponding standard deviation.
The measured sizes are in good agreement with each other and with the expected beam dimension of $460 \pm 20$ nm.
The results support the conclusion that by applying this analysis
we can reconstruct the beam size from the profiles acquired with all three wires. 
It should be noted though that this analysis is only valid for Gaussian beams. 
Conversely, the 1 $\mathrm{\mu m}$ stripe can be used to accurately measure the profile of beams with transverse size of 400-500 nm and directly extrapolate the beam dimension, without the need of an analysis demanding a-priori assumptions on the beam shape. 
Since the resolution limit is lower than the beam dimension the effects of the convolution in the measured profile are negligible, whereas small details would be washed out by the convolution with a larger wire. 
The $\sigma_{y}$ and $\sigma_{\mathrm{rms}}$ of the profile measured with the 1 $\mathrm{\mu m}$ stripe are indeed in good agreement, with an error below 9$\%$. 

\begin{table}[h!]
	\begin{center}
		\begin{tabular}{ r  c  c  c }
			\hline \hline
			&5 $\mathrm{\mu m}$ W & 2 $\mathrm{\mu m}$ Au & 1 $\mathrm{\mu m}$ Au\\ \hline \vspace{.02cm}
			Resolution (nm) & 1250 & 600 &300 \\ \hline\vspace{.02cm}
			$\sigma_{\mathrm{rms}} \; \mathrm{(nm)}$ & 1967 $\pm$ 16&890 $\pm$ 2 &449 $\pm$ 32\\ \hline\vspace{.02cm}
			$\sigma_{{y}} \; \mathrm{(nm)}$ & 462 $\pm$ 11&491 $\pm$ 4 &491 $\pm$ 5\\ \hline\vspace{.02cm}

		\end{tabular}
		\caption{\textbf{Beam vertical size and wire resolution.} The first row shows the wire resolution.
		$\sigma_{\mathrm{rms}}$ is the average rms beam size over three beam profile measurements. $\sigma_{{y}}$ is the mean value of the sigmas from the fit of the three profiles. 
		\label{table:0} }
	\end{center}
\end{table}

\section{Discussion}\label{sec:Discussion}

We have demonstrated the generation of sub-micrometer relativistic electron beams at SwissFEL and their characterization via wire-scanner profile measurements.

This is a particularly promising result in view of advanced accelerator concepts experiments. A notable example is the development of an all-on-a-chip dielectric laser accelerator, pursued by the Accelerator on a Chip International Program \cite{wootton2017towards}. A proof-of-principle experiment of high gradient dielectric laser acceleration will be conducted at PSI using the SwissFEL high-brightness electron beam \cite{ferrari2017achip, prat2017outline}. 
Generating electron beams with sub-micrometer spot size will be mandatory in order to achieve full transmission through the accelerating structure \cite{wootton2016dielectric}.

The characterization of sub-micrometer beams require the development of diagnostics with unprecedented spatial resolution.
Several different techniques are commonly used to measure the beam transverse profile.
Screens are the most used two-dimensional beam profile monitors. 
In the case of the SwissFEL YAG:Ce screens the spatial resolution is 8 $\mathrm{\mu m}$ with a smallest measured beam size of 15 $\mathrm{\mu m}$ \cite{ischebeck2015transverse}.
A higher spatial resolution has been achieved at the UCLA Pegasus laboratory utilizing a 20 $\mathrm{\mu m}$ YAG:Ce crystal with an in-vacuum infinity-corrected microscope objective coupled to a CCD camera. This set-up allowed measurements of beam sizes down to 5 $\mathrm{\mu m}$ and transverse emittances of 5 nm $\times$ 10 nm \cite{maxson2017direct}. 
The spatial resolution of optical transition radiation screens (OTRs) is only limited by the camera sensor and optics 
\cite{tenenbaum1999measurement}.
In the Accelerator Test Facility 2 at KEK, a vertical beam size of 750 nm has been measured by means of OTR monitors \cite{bolzon2015very,kruchinin2014sub}. 
Nevertheless, OTRs are limited by the emission of coherent optical transition radiation from compressed bunches, which prevents their usage in FELs \cite{akre2008commissioning}.
Laser wire scanners are one-dimensional profile monitors that require a laser beam focused to a diffraction-limited waist. The minimum achievable laser waist and therefore measurable beam size is limited by the laser wavelength. At the SLAC Final Focus Test Beam, scientists have measured an electron beam vertical size of 70 nm by means of a laser beam split and folded onto itself to produce an interference fringe pattern \cite{balakin1995focusing}.
The minimum achievable laser wavelength coupled with space constraints and costs represents the main limitation of such profile monitors \cite{tenenbaum1999measurement}. 

In this work we focused on wire scanners, since they are the most promising profile monitors in terms of spatial resolution. 
The latter is limited by the encoder readout, the wire diameter and vibrations. In SwissFEL, a wire-geometry dominated resolution of 1.25 $\mathrm{\mu m}$ was reached with a 5 $\mathrm{\mu m}$ tungsten wire. This was achieved thanks to a motion system provided of an encoder with resolution of 0.1 $\mathrm{\mu m}$ and measured wire-vibrations largely below the geometrical resolution \cite{orlandi2016design, orlandi2017commissioning}.  
Therefore, a natural way to push the geometrical resolution of a WSC towards the sub-micrometer scale was by decreasing the wire diameter.
However, the strength of a wire reduces with its diameter. Thus, the conventional manufacturing technique of stretching a wire onto a wire fork limits its width to a few micrometers \cite{tenenbaum1999measurement}.
We were able to overcome this limitation and fabricate wires with a width down to 1 $\mathrm{\mu m}$ employing innovative nanofabrication techniques.

\section{Methods}\label{sec:measurement}
\subsection{Accelerator setup.}

Our goal was the generation of electron beams with rms vertical size below one micrometer at the interaction point with the wire scanner.  
The rms transverse size of the beam at position $s$ along the accelerator is 
\begin{equation}
\label{eq:BeamDimenison}
\sigma (s) = \sqrt{\frac{\epsilon_n} {\gamma}\;  \beta (s)}\, ,
\end{equation} 
where $\beta(s)$ is the $\beta$-function of the magnetic lattice at position $s$, $\epsilon_n$ is the normalized emittance and $\gamma$ is the electron Lorentz factor.
The beam size can be minimized by reducing the normalized emittance, increasing the beam energy, and reducing the $\beta$-function at the interaction point. 
The final energy $E$ of the SwissFEL injector is 330 MeV, so we focused our efforts in minimizing the emittance and the $\beta$-function at the WSC location.

In RF photoinjectors, the emittance is determined by three different contributions: intrinsic or thermal emittance, space charge, and RF fields \cite{fraser1985high}. To minimize the space-charge contribution to the emittance increase, we operated at a very low charge, below 1 pC. The RF gun was set to its maximum accelerating gradient, while the gun phase was empirically adjusted to minimize the beam energy spread. In these conditions, the emittance was practically determined by the intrinsic or thermal contribution, which is proportional to the laser beam size on the cathode. We therefore reduced the laser iris to the smallest possible diameter, corresponding to an rms laser spot size on the cathode of about 80 $\mathrm{\mu}$m. The electron beam emittance was measured in both transverse planes downstream of the bunch compressor using the symmetric single-quadrupole-scan technique \cite{prat2014symmetric}. The measured normalized emittances are $\epsilon_{n,x}$ = 42 nm and $\epsilon_{n,y}$ = 53 nm in the horizontal and vertical planes respectively, with estimated uncertainties of $\sim 10 \%$. The different transverse emittances derive from a slight asymmetry of the laser spot on the cathode.

Concerning the optics, the strengths of five quadrupole magnets between the bunch compressor and the chamber were calculated to minimize the vertical $\beta$-function at the interaction point, while keeping the horizontal $\beta$-function to a reasonable value. This optimization was performed using the code \ELEGANT \cite{borland2000elegant}. 
Figure {\ref{fig:beta_functions} shows the evolution of the $\beta$-functions from the bunch compressor to the interaction point for the calculated optics. During the experiment, the optics at the exit of the bunch compressor were iteratively matched to their design values using several quadrupole magnets upstream of the bunch compressor. Then, the five quadrupole magnets were set to the pre-calculated settings. Finally, the last quadrupole magnet upstream of the chamber was empirically adjusted to minimize the beam size at the wire-scanner. This adjustment compensated any possible errors in the incoming beam optics at the bunch compressor, the electron beam energy, the quadrupole field, and the exact position of the wire scanner. 

The computed $\beta$-functions at the chamber location are $\beta_{x}$ = 0.273 m in the horizontal plane and $\beta_{y}$ = 2.61 mm in the vertical one. 
According to Eq. ({\ref{eq:BeamDimenison}), the rms vertical beam size is $460 \pm 20$ nm for the considered experimental parameters $E$, $\epsilon_{n,y}$ and $\beta_{y}$ (cf. Table \ref{table:Beam_par}).  
	\begin{figure}[h]
		\centering
		\includegraphics[width = 1\columnwidth]{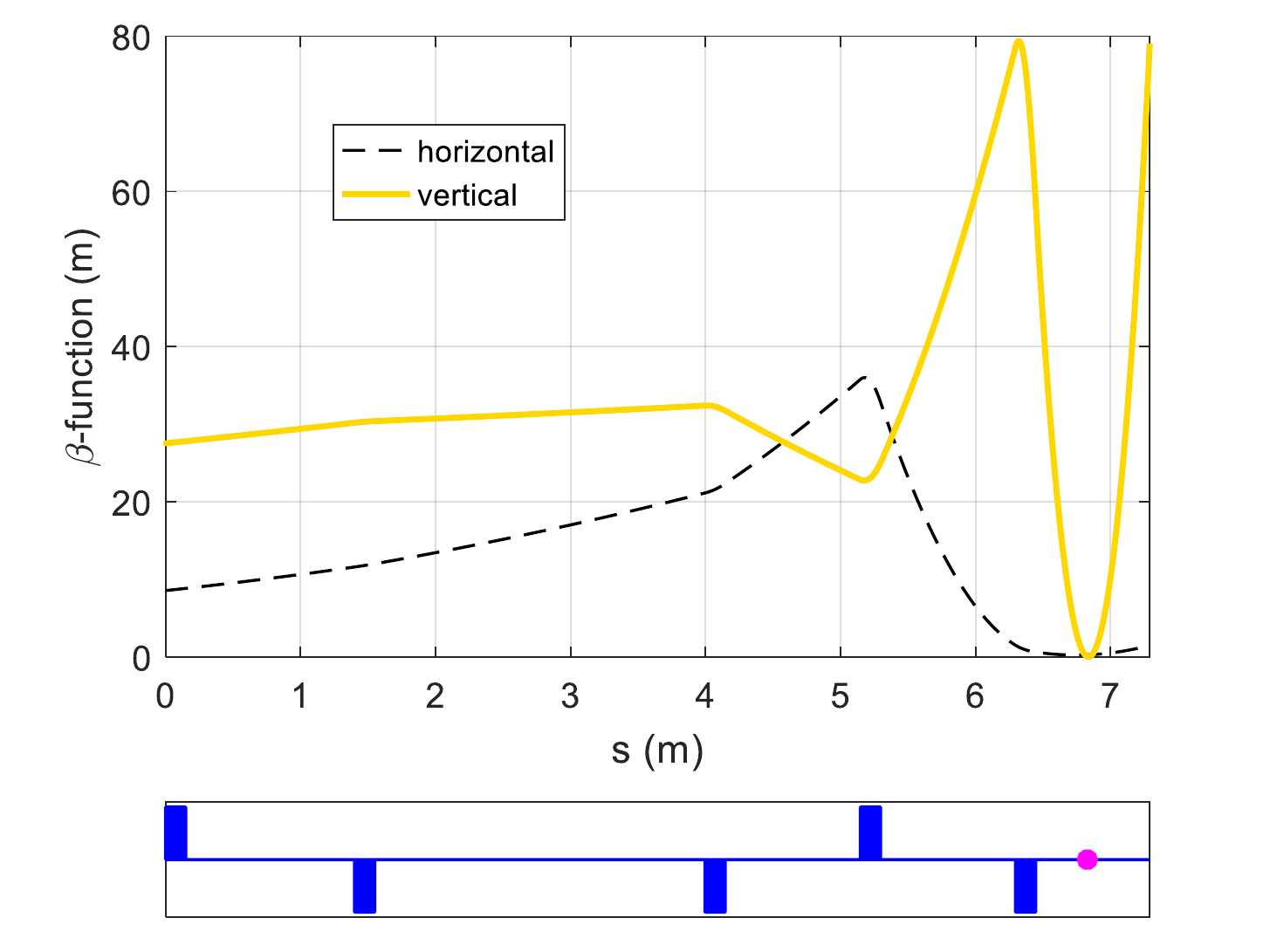}
		\caption{\textbf{Evolution of the $\beta$-functions.} Horizontal and vertical $\beta$-function evolution between the bunch compressor and the experimental chamber location at the SwissFEL injector. The blue rectangles and the magenta dot in the lower plot indicate the location of the quadrupole magnets and the wire scanner, respectively. The magnets are set to obtain a vertical $\beta$-function of 2.61 mm at the wire scanner position. \label{fig:beta_functions}}
	\end{figure}

\subsection{Design and fabrication of sub-micrometer resolution wire-scanner on-a-chip.}
\begin{figure*}
	\centering
	\subfigure[]{
		\includegraphics[height = 0.7\columnwidth]{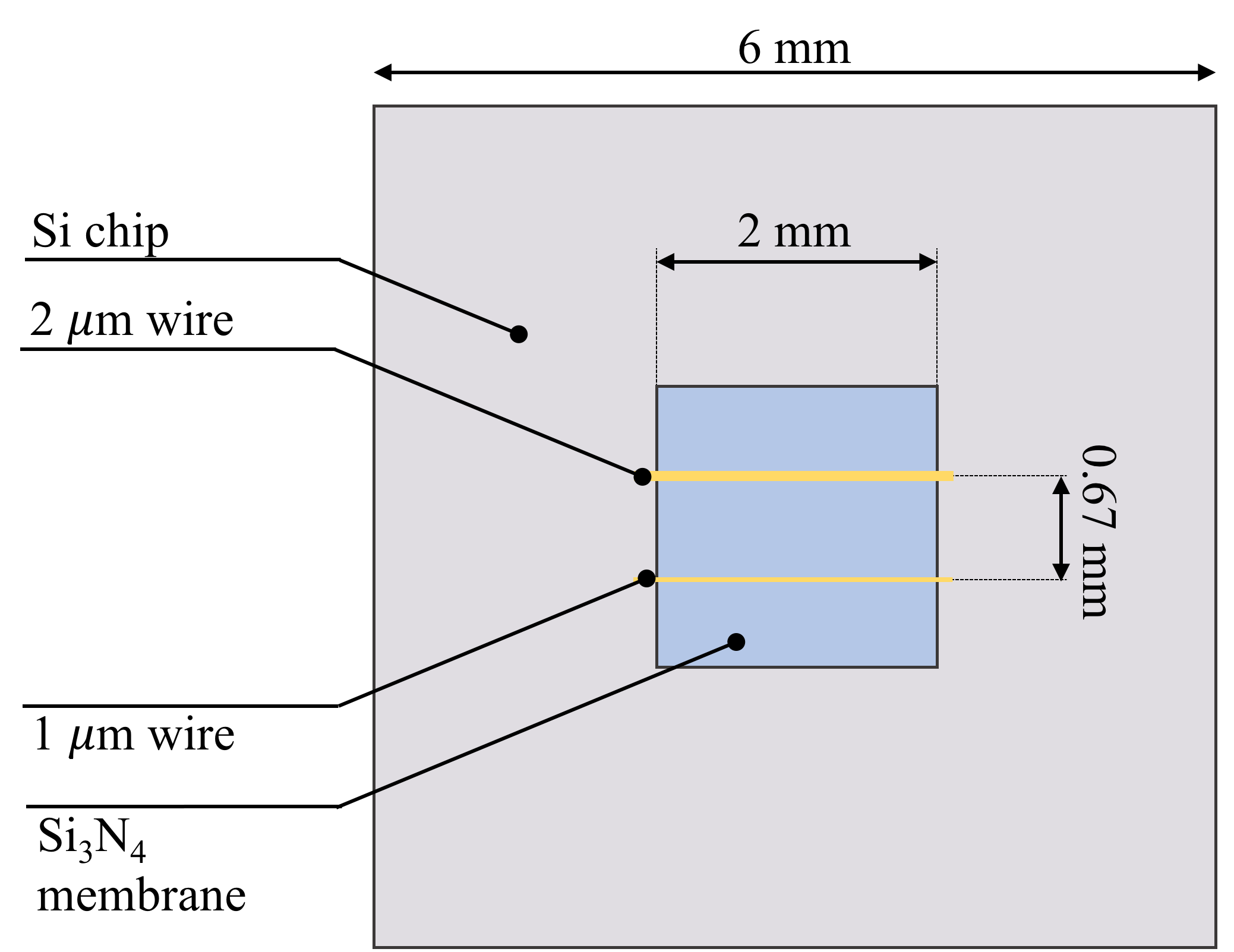}{\label{fig:chip_sch}}
	}
	\subfigure[]{
		\includegraphics[height = 0.622\columnwidth]{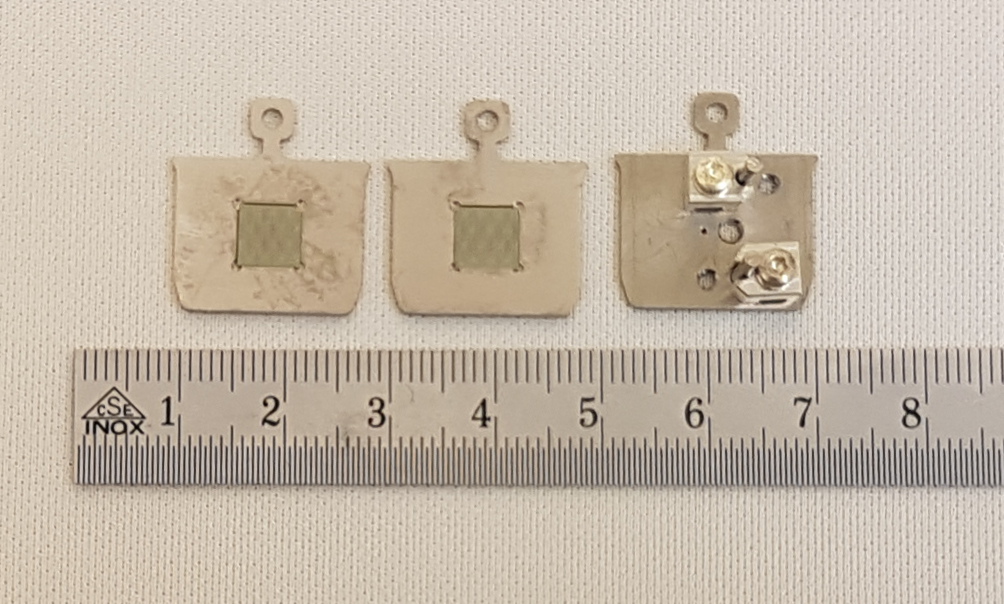}{\label{fig:all_prototypes}}
	}
	\caption{\textbf{Sub-micrometer resolution wire scanners on-a-chip.} \textbf{Panel (a)} Sketch of a sub-$\mathrm{\mu m}$ resolution WSC on-a-chip: 6x6 $\mathrm{mm^2}$ Si chip with a central 2x2 $\mathrm{mm^2}$ $\mathrm{Si_3 N_4}$ 250 nm thick membrane. Two gold stripes of widths $w_1 = 1$ $\mathrm{\mu m}$ and $w_2 = 2$ $\mathrm{\mu m}$ are electroplated on the membrane. The stripes thickness is 3 $\mathrm{\mu m}$. \textbf{Panel (b)} Sub-micrometer resolution wire scanners on-a-chip (center and left) and a conventional 5 $\mathrm{\mu m}$ tungsten wire scanner (right), mounted on sample holders.{\label{fig:wsc_on_a_chip}}}
\end{figure*}
Figure \ref{fig:chip_sch} shows the sketch of a sub-$\mathrm{\mu}$m resolution WSC on-a-chip.
It consists of a 6x6 $\mathrm{mm^2}$ Si chip having a 2x2 $\mathrm{mm^2}$ silicon nitride ($\mathrm{Si_3 N_4}$) membrane at its center. On the membrane two metal stripes of different widths $w_1 = 1 \; \mathrm{\mu m}$ and $w_2 = 2 \; \mathrm{\mu m}$, are electroplated. The separation between the two stripes is 0.67 mm. This distance has been chosen such that each stripe can be separately scanned through the beam.
The stripe material selected is gold. This heavy metal provides a strong scattering cross section for a good signal-to-noise ratio of the loss signal, as well as sufficient resistance against heating from the beam interaction.  

The implemented fabrication procedure is an application of the nanofabrication technique described in \cite{gorelick2010direct, gorelick2011high}.
The prototype fabrication starts with the evaporation of a Cr-Au-Cr metal layer sequence (5-20-5 nm) on a 250 nm thick $\mathrm{Si_3 N_4}$ membrane. 
Then a 3.2 $\mathrm{\mu m}$ thick layer of poly-methyl-methacrylate (PMMA) resist is spin-coated (cf. {Fig. \ref{fig:Frame}). The two Cr layers ensure better adhesion between Au and the membrane as well as the resist \cite{gorelick2010direct}. 
In a second step, applying electron beam lithography we wrote two parallel stripes into the resist, exposing it to a 100 keV electron beam at 5 nA beam current and a dose of 1900 $\mathrm{\mu C/cm^2}$. The electron beam changes the chemical and physical properties of the PMMA resist, making it soluble in the developer solution.
In particular, the exposed regions were developed by immersing the resist in a solution of isopropanol and water (7:3 by volume) and then rinsing the sample in de-ionized water and blow-drying it in a $\mathrm{N_2}$ gas jet.
The development time was 45 s, and as a result the exposed parts of the membrane have been completely removed (cf. {Fig. \ref{fig:Mold}).
After the development, we etched the sample by $\mathrm{Cl_2/CO_2}$ plasma for 45 s, to remove the top Cr layer and reveal the underlying Au sheet.
The developed resist trenches were subsequently filled with Au by electroplating in a cyanide-based plating bath at a plating current density of 2.5 $\mathrm{mA}\;\mathrm{cm^{-2}}$ for 14 minutes.
After the electroplating, the PMMA resist was completely removed by ashing in an oxygen plasma for 15 minutes. In this way, only the Au stripes remained on top of the membrane (cf. Fig. \ref{fig:FinalPrototype}).
The manufactured samples were inspected and characterized with a Scanning Electron Microscope (SEM). Figure \ref{fig:Au_2} shows SEM images of a 2 $\mathrm{\mu}$m and a 1 $\mathrm{\mu}$m wide Au stripe on the $\mathrm{Si_3 N_4}$ membrane. The measured stripe thickness is 3 $\mathrm{\mu}$m.
\begin{figure}
	\centering
	\subfigure[]{
		\includegraphics[height =0.325\columnwidth]{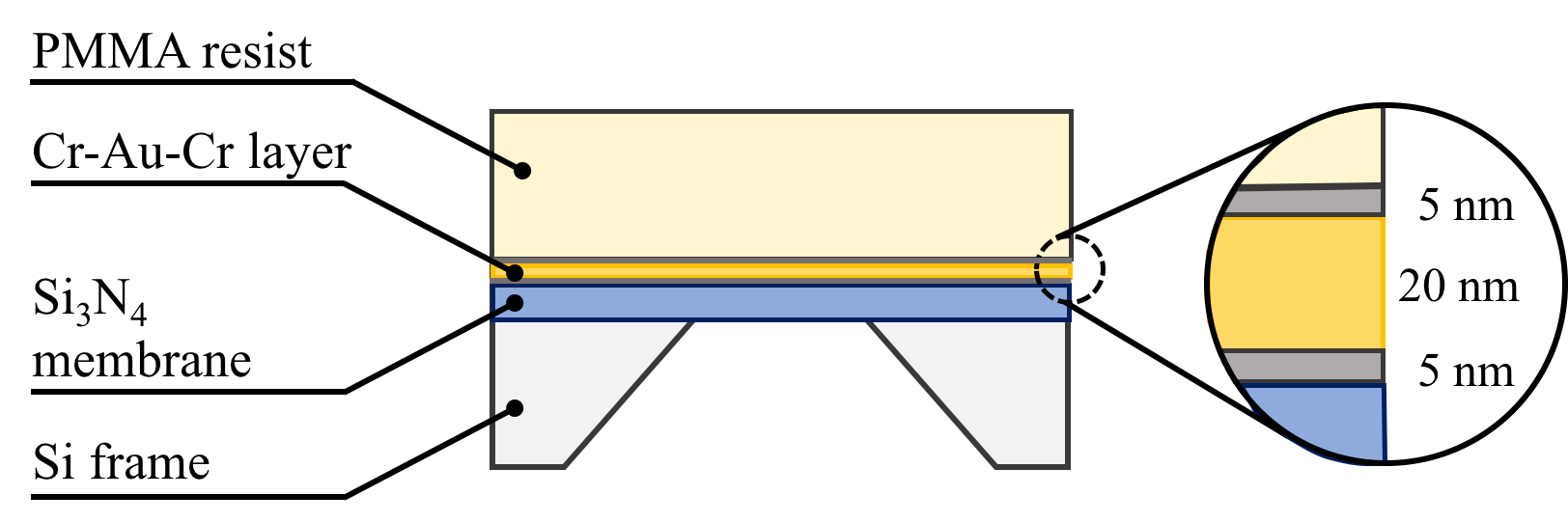}\label{fig:Frame}
	} \,
	\subfigure[]{
		\includegraphics[height = 0.235\columnwidth]{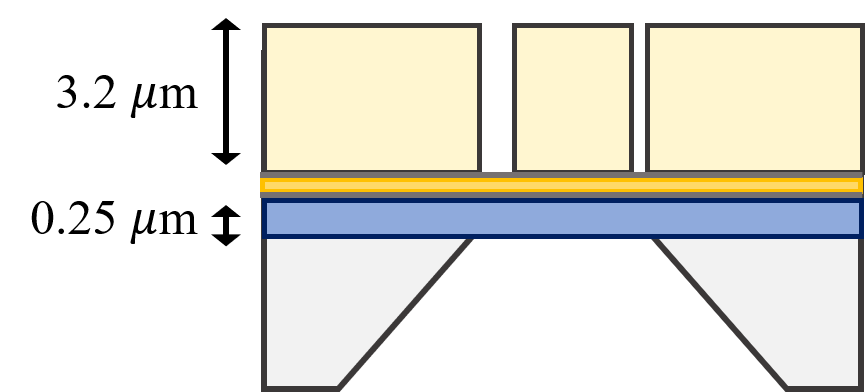}\label{fig:Mold}
	} \,
	\subfigure[]{
		\includegraphics[height = 0.225\columnwidth]{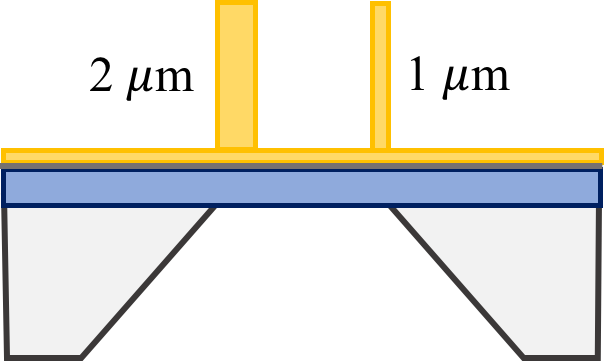}\label{fig:FinalPrototype}
	} \\
	\caption{\textbf{Block diagram of the WSC on-a-chip fabrication process} (not to scale). \textbf{Panel (a)} $\mathrm{Si_3 N_4}$ membrane deposited on a Si chip. The membrane is coated with a Cr-Au-Cr layer. A 3.2 $\mathrm{\mu}$m thick layer of PMMA resist is spin-coated on this Cr-Au-Cr layer. \textbf{Panel (b)} The PMMA resist is exposed by electron beam lithography to write in it two parallel stripes. Exposed resist regions are then developed by immersion in a mixture of isopropanol and water. \textbf{Panel (c)} The developed membrane trenches are filled with Au by electroplating. The PMMA resist is removed by ashing in an $\mathrm{O_2}$ plasma. \label{fig:FabricationProcess}}
\end{figure}

\begin{table*}[]
	\begin{center}
		\newcolumntype{R}{>{\centering\arraybackslash}X}%
		\begin{tabularx}{0.85\textwidth}{R R R R R R } 
			\hline 
			\multirow{2}{*}{$E$ (MeV)} & \multirow{2}{*}{$Q$ (pC)} & \multicolumn{2}{c}{$\beta$-functions}& \multicolumn{2}{c}{Emittances} \\
			\cline{3-6}
			& & $\beta_x$ (m) &$\beta_y$ (mm) & $\epsilon_{n,x}$ (nm) & $\epsilon_{n,y}$ (nm) \\ \hline 
			330 &  $<$ 1 & 0.273 & 2.61 & 42 & 53 \\
			\hline
		\end{tabularx}
		
		\caption{\textbf{Beam parameters.} First and second column show the electron energy and the beam charge. 
			$\epsilon_{n,x}$ and $\epsilon_{n,y}$ are the measured normalized emittances in the vertical and horizontal plane. $\beta_{x}$ and $\beta_{y}$  are the computed horizontal and vertical $\beta$-function at the interaction point.
			\label{table:Beam_par} }
	\end{center}
\end{table*}

\begin{figure}[]
	\centering
	\subfigure[]{
		\includegraphics[height = 0.55\columnwidth,angle =0]{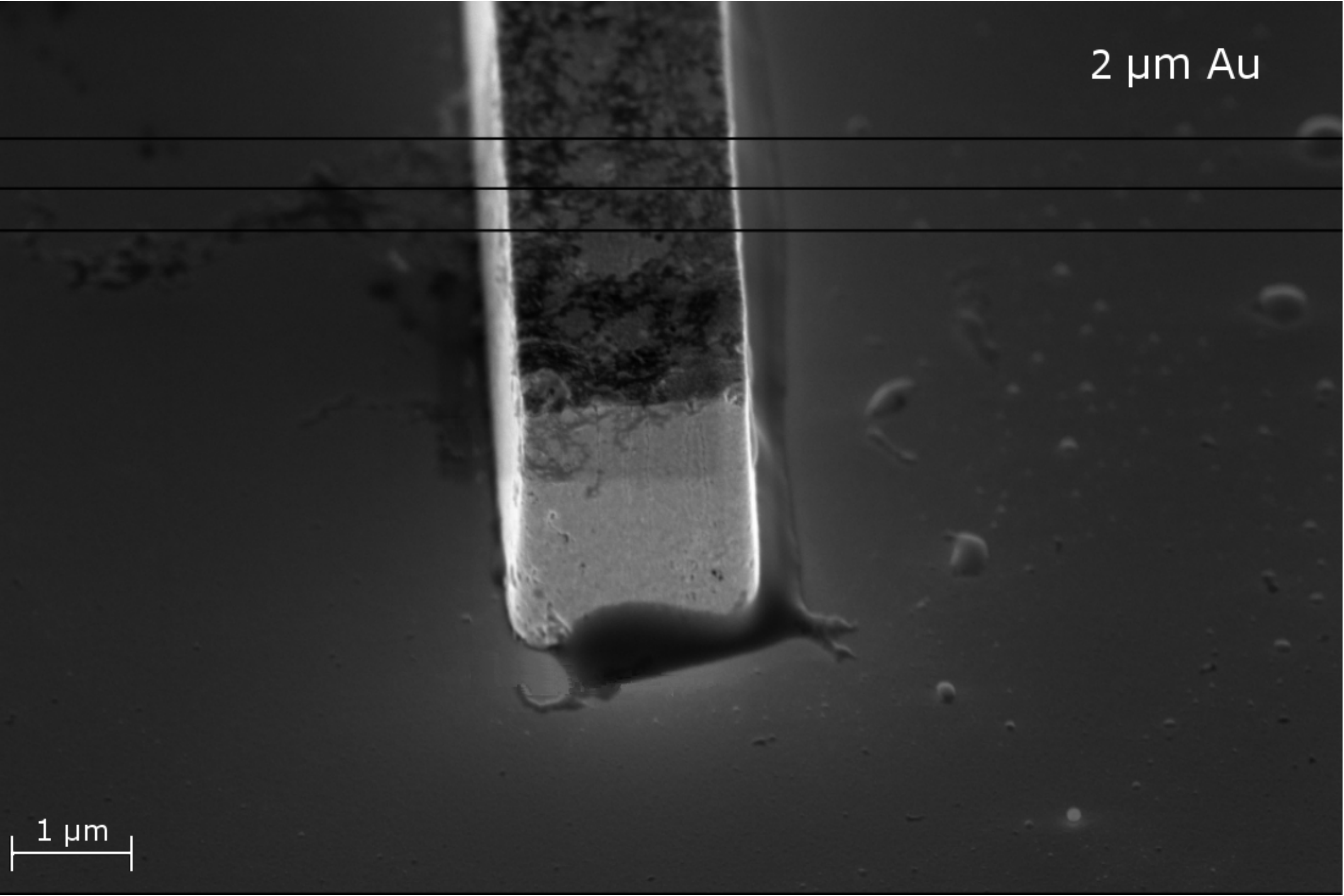}
	}\,
	\subfigure[]{
		\includegraphics[height = 0.55\columnwidth,angle =0]{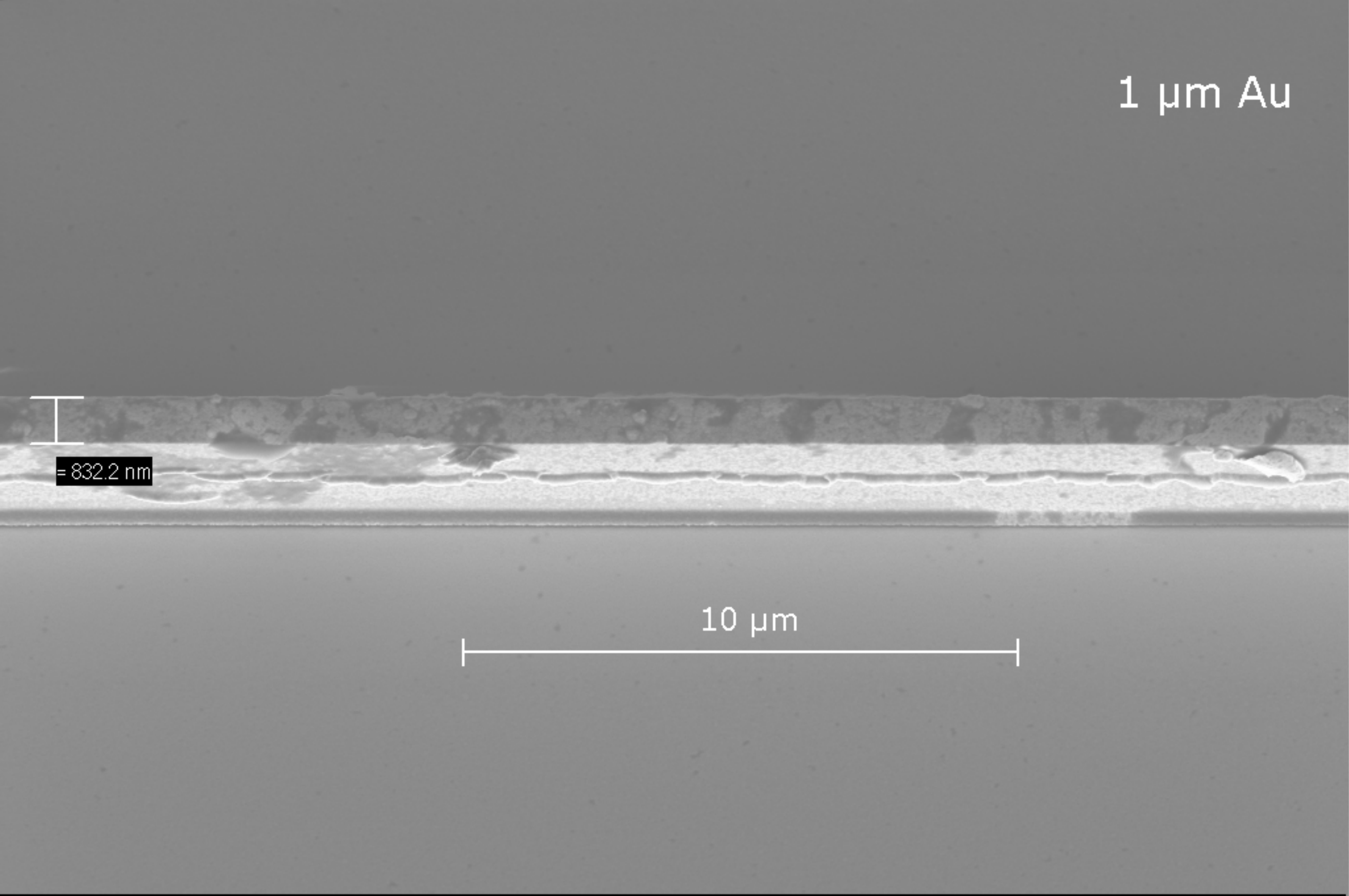}
	}\\
	\caption{\textbf{Scanning electron microscope pictures of nanofabricated stripes.} The gold stripes are electroplated onto a $\mathrm{Si_{3}N_4}$ membrane. \textbf{Panel (a)} Tilted view of the 2 $\mathrm{\mu}$m wide gold stripe edge. The viewing angle is 50$^{\circ}$. The stripe thickness was measured from the SEM image to be 3 $\mathrm{\mu}$m. \textbf{Panel (b)} Tilted view of the 1 $\mathrm{\mu}$m wide gold stripe.\label{fig:Au_2}}
\end{figure}

\subsection{Detection setup.}
Figure \ref{fig:ChamberSchematic} shows a schematic drawing of the experimental chamber in the SwissFEL injector, employed to perform the wire scanner measurements. The chamber is equipped with an in-vacuum manipulator operated by a 2-phase stepper motor via a feed-through. Four sample holders are mounted on the manipulator where four different samples can be settled. The stepper motor translates vertically the manipulator so that a sample can be inserted into the SwissFEL vacuum chamber along this direction and brought to the interaction point with the electron beam.
By using a load-lock pre-chamber, the samples can be installed on the sample holders without breaking the vacuum employing a manually controllable manipulator. \cite{ferrari2017achip}.  
 \begin{figure}
	\centering
	\subfigure[]{
		\includegraphics[width = 0.425\textwidth]{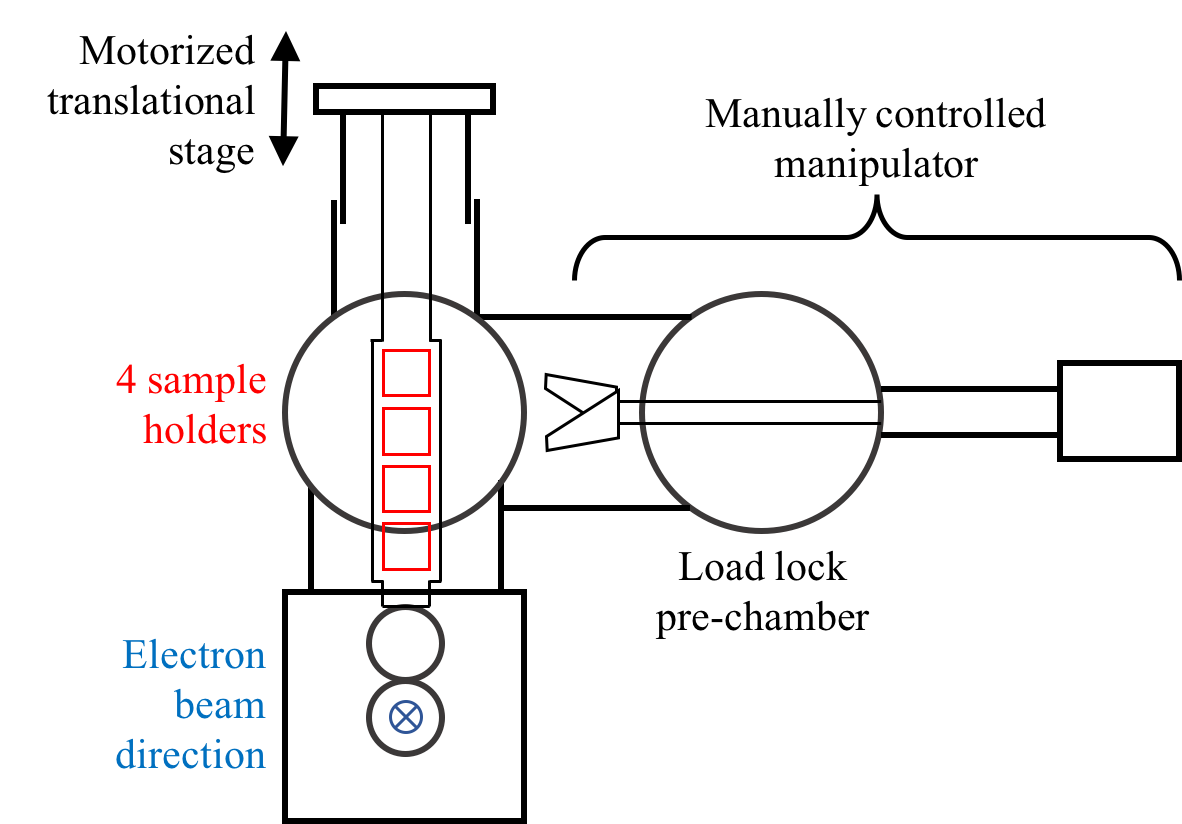}\label{fig:ChamberSchematic}
	}
	\subfigure[]{
		\includegraphics[width = 0.325\textwidth]{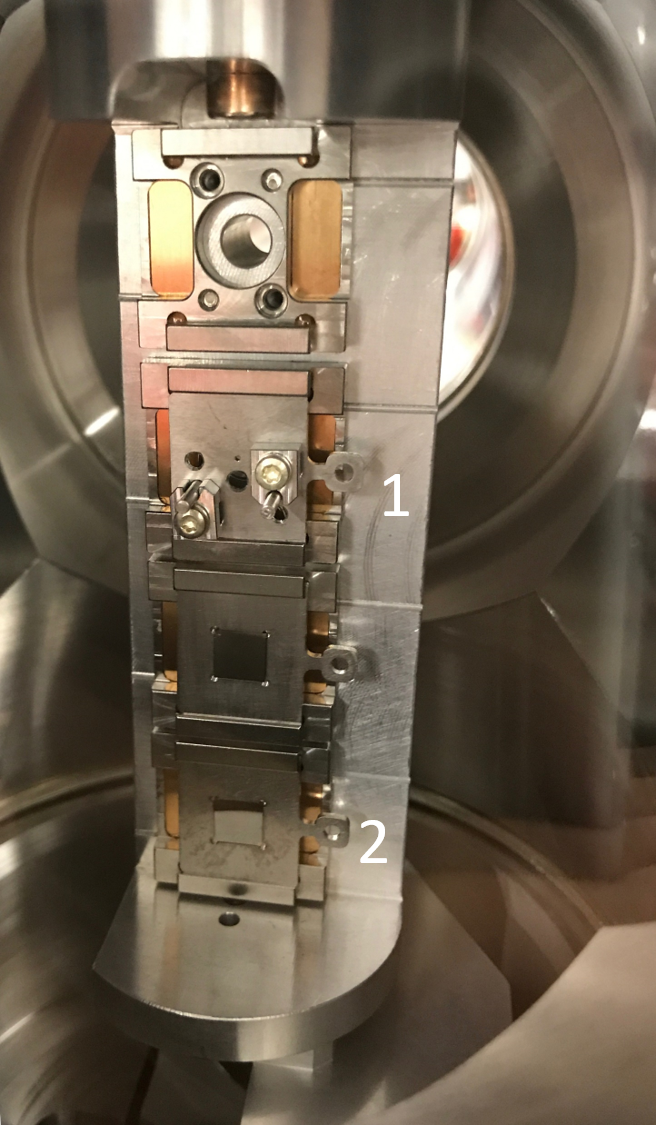}\label{fig:PrototypeInstalled}
	}
	\caption{\textbf{SwissFEL Injector experimental chamber and WSC prototypes installed in the beamline.} \textbf{Panel (a)} Schematic drawing of the experimental chamber. An in-vacuum manipulator, operated by a 2-phase stepper motor, is equipped with four different sample holders to harbor samples. A load-lock pre-chamber allows the insertion of samples in the beam pipe without breaking the accelerator vacuum using a manually controllable manipulator. \textbf{Panel (b)} Wire scanners installed in the SwissFEL Injector beamline: (1) conventional 5 $\mathrm{\mu}$m tungsten wire, (2) wire scanner on-a-chip.
	}
\end{figure}

In the chamber we installed the sub-$\mathrm{\mu}$m resolution WSC on-a-chip as well as a reference WSC consisting of a 5 $\mathrm{\mu}$m tungsten wire. The wire in this case is mounted using the conventional technique of stretching it onto the sample holder and fixing it between two pins (cf. Fig. \ref{fig:all_prototypes}). 
{Fig. \ref{fig:PrototypeInstalled} shows the prototypes installed in the SwissFEL injector beamline.

During a WSC beam profile measurement, the wire is scanned through the beam along the vertical direction at a constant velocity. The wire position is acquired by an absolute optical encoder (0.1 $\mathrm{\mu}$m resolution).
When the electron beam intercepts the wire a shower of primary scattered electrons and secondary emitted $e^+$ $e^-$ $\gamma$ is produced. The shower intensity is proportional to the fraction of the beam intercepted by the wire and represents the signal of interest (loss signal).
The loss signal is detected outside of the vacuum chamber by a beam loss monitor (BLM) installed 4.5 m downstream of the experimental chamber. SwissFEL BLMs consist of a scintillator fiber (Saint Gobain BCF-20) wrapped around the beam pipe. This scintillator fiber is connected via a plastic optical fiber to a photomultiplier tube, which has remotely adjustable gain in the range $[5 \cdot 10^3,\, 4 \cdot 10^6]$. The signal is finally sent to an ADC unit for digitization and processing \cite{Pollet:IBIC2015-MOPB051}.
The BLM gain was set to $9 \cdot 10^4$. This value guaranteed a suitable signal quality and amplification during the beam profile measurement with the three wires.

For every wire position along the beam we measured the loss signal for one second at a beam repetition rate of 10 Hz. 
Every point in the beam profiles presented in Fig. \ref{fig:BeamProfiles} is the average of these acquired values. 

\subsection{Data processing.}
 
Every infinitesimal element of the wire width induces on the beam a loss that is proportional to the product of the wire thickness and the beam distribution \cite{Fernow}. 
Under the assumption of a Gaussian beam shape the measured loss signal is given by 
\begin{equation}
\label{eq:convolutionFunction}
 f(y;\, \Delta,\alpha,\sigma,\gamma)= \int_{-\infty}^{\infty}t(u) \, \left[\Delta + \alpha \,e^{\frac{-(u-y-\gamma)^{2}}{2 \, \sigma^{2}}}\right] \;\mathrm{d}u \, , 
\end{equation}
where $y$ is the wire position along the measurement direction and $t(\cdot)$ is a function describing the wire thickness shape. 
$\alpha$,\,$\gamma$ are the Gaussian function amplitude and centroid respectively; $\Delta$ is an offset and $\sigma$ is the Gaussian function standard deviation.   
The data were fitted through this function, with free parameters $\Delta, \alpha, \sigma, \gamma$.
 
The conventional 5$\mu$m tungsten wire is cylindrical, hence to describe the wire shape we choose the function 
\begin{equation}
\label{eq:cylindrical}
t(u) =
\begin{cases}
2\sqrt{\left({\frac{D}{2}}\right)^{2}-u^{2}} \quad &\mathrm{for} \;  u \in \left[-\frac{D}{2},\frac{D}{2}\right]\\
0 \quad &\mathrm{otherwise} \, ,
\end{cases} 
\end{equation} 
where $D = 5 \;\mathrm{\mu m}$ is the wire diameter.

Instead, the 2 and 1 $\mathrm{\mu m}$ gold stripes are rectangular so the proper function to describe their shape is the step function
\begin{equation}
\label{eq:step}
t(u) =
\begin{cases}
t_w \quad &\mathrm{for} \;  u \in \left[-\frac{w}{2},\frac{w}{2}\right]\\
0 \quad &\mathrm{otherwise}\, ,
\end{cases} 
\end{equation}  
where $t_w$ = 3 $ \mathrm{\mu m}$ is the stripe thickness and $\textit{w}$ is the stripe width.

Since the tungsten wire radius and the widths of the gold stripes are
known from measurements, it is not necessary to consider them free parameters in the fit.

\subsection{Wire scanner geometrical resolution.}
The geometrical resolution $r$ of a wire is defined by its rms size \cite{tenenbaum1999measurement, orlandi2016design}.
The square rms size of a wire whose thickness is described by the function $t(\cdot)$ is
\begin{equation}
\sigma_{rms}^2= \frac{\int_{-\infty}^{+\infty}t(u) \, (u-<u>)^2\;\mathrm{d}u }{\int_{-\infty}^{+\infty}t(u) \;\mathrm{d}u }\, ,
\end{equation}
where
\begin{equation}
<u>=\frac{\int_{-\infty}^{+\infty}t(u) \, u\;\mathrm{d}u }{\int_{-\infty}^{+\infty}t(u) \;\mathrm{d}u }\, .
\end{equation}
A wire that is round in cross-section has an rms size equal to ${D}/{4}$ \cite{field1995wire}, since its thickness is described by the function $t(u)$ in Eq. \ref{eq:cylindrical}. As a result, the 5 $\mathrm{\mu m}$ W wire geometrical resolution is equal to  $r$ = 1.25 $\mathrm{\mu m}$.
For a rectangular stripe we have $ r = {w}/{\sqrt{12}}$ (cf. Eq. \ref{eq:step}).
Therefore, the geometrical resolutions of the 1 and 2 $\mathrm{\mu m}$ Au stripes are equal to 0.3 and 0.6 $\mathrm{\mu}$m, respectively.
 
\subsection{Data availability.}
The data that support the findings of this study are available
from the corresponding author upon request.

\subsection{Competing Interests.}
The authors declare that they have no competing financial interests.

\bibliographystyle{naturemag} 
\bibliography{paper_biblio}

\section{Acknowledgements}
The authors would like to express their sincere thanks to Thomas Schietinger for his careful proof-reading of the manuscript and to  Beat Rippstein for his help during the mounting of the wire scanner on-a-chip. We are furthermore grateful to the SwissFEL vacuum, operation and laser groups for their support.
This research is supported by the Gordon and Betty Moore Foundation through Grant GBMF4744 to Stanford.

\section{Author Contributions}
M.B., S.B., C.D., R.I., G.L.O. designed the wire scanner on-a-chip. V.G. nanofabricated the devices and with S.B. characterized them. S.B., R.I., G.L.O. designed the experimental set-up. 
E. P. calculated the optics and minimized the beam size at the wire-scanner location during the experiment.
E. F. and E. P. set up and transported the electron beam through the accelerator.
S.B., E.F. and R.I. installed the wire scanners. 
R.I. and E.P. wrote the data acquisition codes.
S.B., E.F., R.I., G.L.O., C.O.L., E.P. performed the beam profile measurements.
S.B., R.I. and G.L.O. performed the data analysis.
S.B. wrote the manuscript, which was enriched by all the co-authors suggestions.

\end{document}